\documentclass[prd,twocolumn,amsmath,amssymb,showpacs,floatfix,nofootinbib,preprintnumbers,superscriptaddress]{revtex4-1}
\usepackage[latin1]{inputenc}
\usepackage[normalem]{ulem}
\usepackage{amsmath}
\usepackage{amsfonts}
\usepackage{amssymb}
\usepackage{graphicx}
\usepackage{dcolumn}
\usepackage{subcaption}
\usepackage{bm}
\usepackage{ulem}
\usepackage{latexsym}
\usepackage{epsfig}
\usepackage{graphicx}
\usepackage[colorlinks=true]{hyperref}
\usepackage[dvipsnames]{xcolor}
\renewcommand{\vec}[1]{\bm{#1}}

\newcommand{\s}{\textcolor{black}}

\begin{document}
\title{Accurate precision Cosmology with redshift unknown gravitational wave sources}
\author{Suvodip Mukherjee}\email{ s.mukherjee@uva.nl, mukherje@iap.fr}
\affiliation{Gravitation Astroparticle Physics Amsterdam (GRAPPA),
Anton Pannekoek Institute for Astronomy and Institute for High-Energy Physics,
University of Amsterdam, Science Park 904, 1090 GL Amsterdam, The Netherlands}
\affiliation{Institute Lorentz, Leiden University, PO Box 9506, Leiden 2300 RA, The Netherlands}
\affiliation{Delta Institute for Theoretical Physics, Science Park 904, 1090 GL Amsterdam, The Netherlands}
\author{Benjamin D. Wandelt}\email{bwandelt@iap.fr}
\affiliation{Institut d'Astrophysique de Paris (IAP), UMR 7095, CNRS/UPMC Universit\'e Paris 6, Sorbonne Universit\'es, 98 bis boulevard Arago, F-75014 Paris, France}
\affiliation{ Institut Lagrange de Paris (ILP), Sorbonne Universit\'es, 98 bis Boulevard Arago, 75014 Paris, France}
\affiliation{Center for Computational Astrophysics, Flatiron Institute, 162 5th Avenue, New York, NY 10010, USA}
\author{Samaya M. Nissanke}\email{samaya.nissanke@uva.nl}
\affiliation{Gravitation Astroparticle Physics Amsterdam (GRAPPA),
Anton Pannekoek Institute for Astronomy and Institute for High-Energy Physics,
University of Amsterdam, Science Park 904, 1090 GL Amsterdam, The Netherlands}
\author{Alessandra Silvestri}\email{silvestri@lorentz.leidenuniv.nl}
\affiliation{Institute Lorentz, Leiden University, PO Box 9506, Leiden 2300 RA, The Netherlands}
\date{\today}
\begin{abstract}
Gravitational waves can provide an accurate measurement of the luminosity distance to the source, but cannot provide the source redshift unless the degeneracy between mass and redshift can be broken. This makes it essential to infer the redshift of the source independently to measure the expansion history of the Universe. We show that by exploiting the clustering scale of the gravitational wave sources with galaxies of known redshift, we can infer the expansion history from redshift unknown gravitational wave sources. By using gravitational wave sources of unknown redshift that are detectable from the network of gravitational wave detectors with Advanced LIGO design sensitivity, we will be able to obtain accurate and precise measurements of the local Hubble constant, the expansion history of the universe, and the gravitational wave bias parameter, which captures the distribution of gravitational wave sources with respect to the redshift tracer distribution. While we showcase its application to low redshift gravitational waves, this technique will be applicable also to the high redshift gravitational wave sources detectable from Laser Interferometer Space Antenna (LISA), Cosmic Explorer (CE), and Einstein Telescope (ET). Moreover, this method will also be applicable to samples of supernovae and fast radio bursts with unknown or photometric redshifts. 

\end{abstract}
\maketitle
\section{Introduction}
\s{The} measurement of the current expansion rate of the Universe, known as Hubble constant (denoted by $H_0$), as well as its value at different cosmological redshifts, is one of the key science goals in the field of \s{c}osmology. This endeavour, which started with the first measurement of $H_0$ by Edwin Hubble \cite{1929PNAS...15..168H} has been typically performed via electromagnetic probes which can be classified as standardized candles (e.g., supernovae (SNe)) \cite{Perlmutter:1998np,2009ApJ...695..287R, Riess:2019cxk}, standard rulers (e.g., cosmic microwave background (CMB), baryon acoustic oscillation (BAO)) \cite{Ade:2013zuv, Anderson:2013zyy, Aubourg:2014yra, Ade:2015xua, Macaulay:2018fxi}, and \s{a} standard clock \cite{Jimenez:2001gg, Simon:2004tf, Stern:2009ep, Jimenez:2019onw}. All these probes have become increasingly successful in making \s{precise} measurements of $H_0$, but have failed to converge to \s{values which are consistent with each other within their error-bars (including both statistical and known systematic uncertainties)}. In fact, low redshift probes such as SNe \cite{Riess:2019cxk} \s{indicate} a value of $H_0= 74 \pm 1.4$ km/s/Mpc , whereas the probes which depend on the high redshift Universe such as big bang nucleosynthesis (BBN), CMB, BAO indicate a value of \s{ $H_0= 67.4 \pm 0.5$ } km/s/Mpc \cite{Abbott:2017smn, Aghanim:2018eyx}. An independent measurement of $H_0= 73. 8^{+1.7}_{-1.8}$ km/s/Mpc from the time delay of the strongly\s{-}lensed low redshift events by the H0LiCOW \cite{Wong:2019kwg} also supports the mismatch. {The discrepancy in the value of $H_0$ between early-time and late-time probes is more than $4\sigma$} \cite{Verde:2019ivm}. {We shall note that independent, late-time measurements that use the Tip of the Red Giant Branch (TRGB) to calibrate SNe have recently given $H_0= 69.8\, \pm \,0.8\, \text{(stat)} \pm 1.7\, \text{(sys)}$ km/s/Mpc \cite{Freedman:2019jwv}, which reduces the discrepancy significantly}. There are also \s{studies that propose possible sources of systematics} in the late-time measurements of $H_0$ \cite{Rigault:2018ffm, Kochanek:2019ruu}. As of yet, there is no conclusive evidence {that} settles this mismatch by {either} any systematic, {and/or} invoking new physics. Independent probes are required to settle this discrepancy.  

The direct detection of gravitational waves has recently offered a new independent probe of cosmic expansion. From the gravitational wave chirp generated by compact object binary mergers, one can infer the luminosity distance of the source \cite{1986Natur.323..310S, Holz:2005df, Dalal:2006qt, PhysRevD.77.043512, 2010ApJ...725..496N, 2011CQGra..28l5023S, Nissanke:2013fka} leading to \s{gravitational wave} sources being dubbed standard sirens. Interestingly the intrinsic luminosity of the gravitational wave source depends on the chirp mass, and its evolution with the \s{frequency is solely dictated by} the general theory of relativity \cite{1986Natur.323..310S, Holz:2005df}, {without the need of external calibration. The only limiting factors are any systematic \s{uncertainties} arising from the gravitational wave detector calibration \cite{Sun:2020wke,Bhattacharjee:2020yxe} and statistical uncertainty arising from the determination of the inclination angle, which is degenerate with the luminosity distance in setting the strain} \cite{2010ApJ...725..496N}. 

\begin{figure*}
\centering
\includegraphics[trim={0.cm 0.cm 1.cm 0.5cm},clip,width=1.\textwidth]{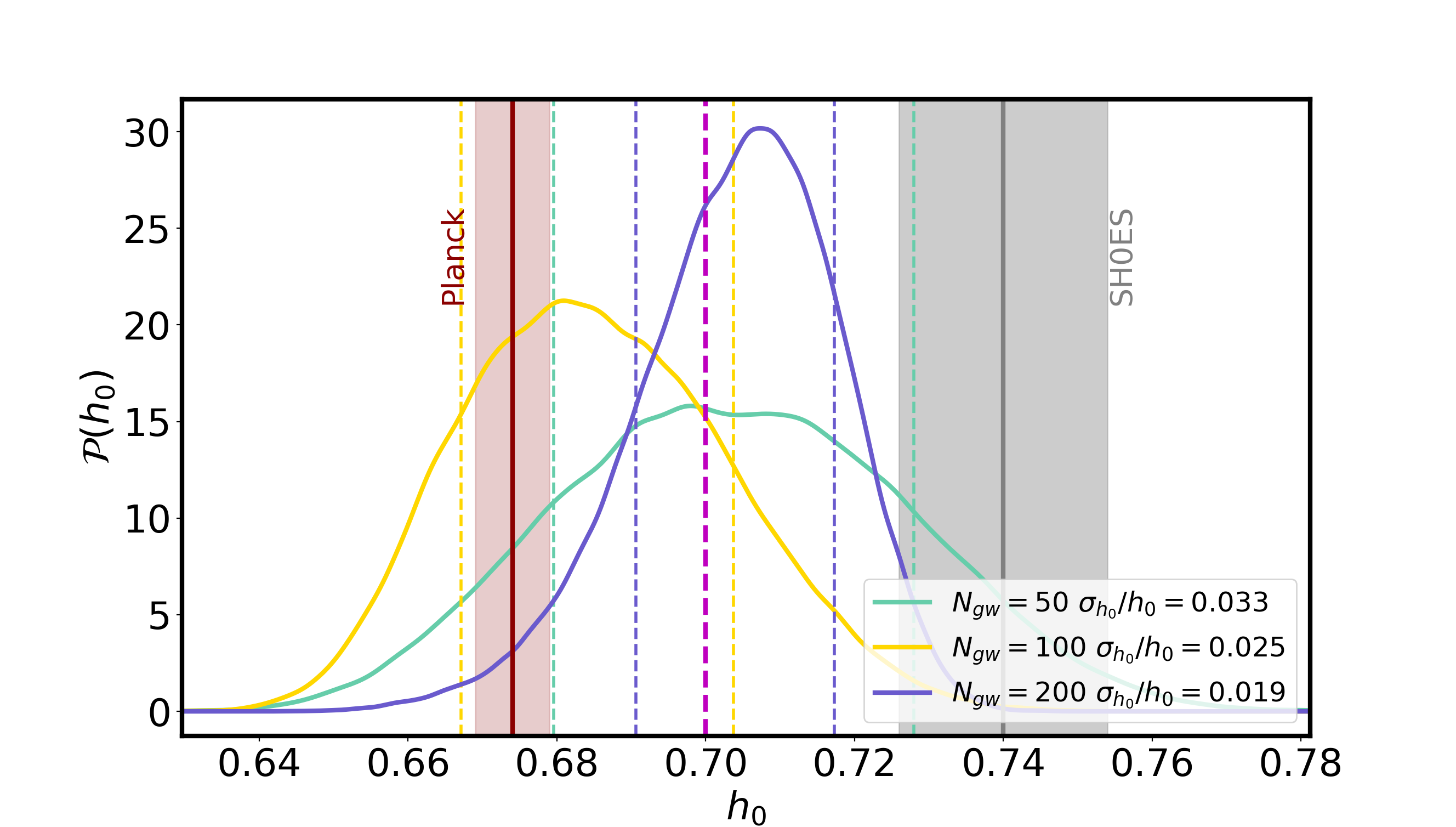}
\captionsetup{singlelinecheck=on,justification=raggedright}
\caption{We show the normalised posterior of the Hubble constant $H_0= 100\,h_0$ km/s/Mpc for different number\s{s} of gravitational wave sources distributed up to redshift $z=0.5$ for the sky localization error $\Delta \Omega_{GW}= 10$ sq. deg\s{.} after marginalizing over cosmological parameter $\Omega_m$ and nuisance parameters related to the gravitational wave bias parameters $b_{GW}(z)$. The constraints are also similar for $\Delta \Omega_{GW}= 25$ sq. deg. In vertical dashed line we show the region between $16^{th}$ and $84^{th}$ percentile of the distribution for each cases. The vertical magenta dashed line denotes the injected value of $h_0=0.7$ indicating reliable recovery in all cases. For comparison we also plot the measured value of $h_0= 0.674\pm 0.005$ by the Planck collaboration \cite{Ade:2015xua} and the value of $h_0=0.74 \pm 0.014$ by the SH0ES Team \cite{Riess:2019cxk}.}
\label{allh0}
\end{figure*}

Though standard sirens are promising, using them for the measurement of the expansion history \s{requires} an independent measurement of their redshift. The gravitational wave signal alone does not provide this information in the absence of a known scale arising from either the tidal deformation \cite{Messenger:2011gi}\s{,} or {the} mass-gap in the binary black hole (BBH) sources due to pair-instability supernova \cite{Farr:2019twy}. Another possibility to determine the redshift is by identifying the host galaxy using a coincident detection of an electromagnetic (EM) counterpart from the gravitational wave source. {The first-ever multi-messenger detection happened for the binary neutron star merger GW170817 detected by the Laser Interferometer Gravitational-wave Observatory (LIGO) Scientific {and} Virgo Collaborations (LVC), for which the GRB170817A electromagnetic counterpart was observed, leading}  to an independent measurement of the Hubble constant $H_0= 70_{-8}^{+12}$ \s{km/s/Mpc} \cite{Abbott:2017smn}. As shown in \cite{Howlett:2019mdh, Mukherjee:2019qmm, Nicolaou:2019cip}, the joint estimation of the electromagnetic signal and gravitational wave signal requires peculiar velocity correction to the gravitational wave sources. In general, the error bar on $H_0$ is more than $15\%$ and is currently not competitive with the measurements from CMB ($<1\%$) and SNe ($\sim 1.5\%$). However in the future\s{, with the} measurement of a large number of sirens ($\sim 50$) with EM counterparts, one can achieve a $2\%$ measurement of $H_0$ \cite{Chen:2017rfc, Feeney:2018mkj}. Another \s{way} to reduce the error-bar on the value of $H_0$ is by measurement of the inclination angle\footnote{The angle between the line of sight and the system's orbital angular momentum.} by either measuring the two polarization states of the gravitational wave signal using an expanded network of three or more gravitational wave detectors \cite{2010ApJ...725..496N}\s{,} {or by using the higher-order multipole moments of the gravitational wave signal \cite{LIGOScientific:2020stg}. Measurement of the inclination angle is also possible }by accurately modeling the EM emission from the jet of the gravitational wave source (e.g., \cite{Ghirlanda:2018uyx, Mooley:2018qfh, Hotokezaka:2018dfi}\s{)}, though this method may introduce astrophysical modeling uncertaint\s{ies}. 

{Consistent with the \s{only definitive} binary neutron star (BNS) detection with \s{an} EM counterpart so far \cite{TheLIGOScientific:2017qsa, Abbott:2017xzu}, the expected number of gravitational wave sources with an EM counterpart in the cosmic volume that can be explored by the Advanced LIGO/Virgo and KAGRA detectors {is expected to be} small since; in fact,  only {a fraction of the total} BNS and neutron star black hole (NS-BH) systems events are expected to have a detectable EM counterpart \cite{Foucart:2018rjc, Coughlin:2019xfb, Andreoni:2019qgh, Abbott:2020uma, Abbott2020}.}
Successful detection of the EM counterpart requires its flux to be higher than the detection threshold of follow-up telescopes. \s{It also requires the sky localization area of the gravitational wave source to be small enough to do a fast search of the EM counterpart before it fades away} \cite{Kasliwal:2020wmy}. As a result, BNS and NS-BH systems\s{,} which are farther away {with} poor sky localization\s{,} may not have a detectable EM counterpart, \s{similar to the possible BNS (or NS-BH)} event \s{GW190425} \cite{Abbott:2020uma}.  
All these issues can be a serious bottleneck for measuring $H_0$ using gravitational wave sources in the \s{timescale} of ten years with \s{a precision} of $\sim 2\%$ \cite{Chen:2017rfc, Feeney:2018mkj, Vitale:2018wlg, PhysRevD.100.103523}. 

\s{Gravitational wave sources\s{,} such as BBHs which have higher intrinsic luminosity \s{that} can be detected \s{at} \s{farther} distances in comparison to BNS {systems}, granting us access to a larger {detectable} cosmic volume}. However, the {majority of} BBHs which are detectable in the {frequency band of} the Advanced LIGO/Virgo detectors are not expected to have an EM counterpart \s{by themselves,} unless there is a presence of baryons surrounding the BBH, {where a candidate was recently announced} \cite{PhysRevLett.124.251102}. We refer to the astrophysical systems without any EM counterparts as \textit{dark standard sirens}. Due to the absence of the EM counterpart, identification of the host galaxy is not possible, and hence their redshift cannot be identified in the standard way. {An alternative approach is required to exploit the expected large number of dark standard sirens for a measurement of the Hubble constant.} 

A possibility is to statistically obtain the host galaxy of dark sirens from galaxy catalogs \cite{DelPozzo:2012zz, Chen:2017rfc, Nair:2018ign, PhysRevD.101.122001}. An application of this for the existing gravitational wave data was performed in previous studies \cite{Fishbach:2018gjp, Abbott:2019yzh, Soares-Santos:2019irc, Abbott:2020khf, Palmese:2020aof}. These methods can be promising but are not optimal, as we will discuss in the following section. Forecast studies of this method report the possibility of making $H_0$ measurement at the level of $5.8\%$ in the future with $50$ objects \cite{Chen:2017rfc, Nair:2018ign, Soares-Santos:2019irc}\footnote{Scaling the previous bounds from \cite{Chen:2017rfc} and \cite{Soares-Santos:2019irc} as $1/\sqrt{N_{GW}}$ indicates similar error-bar.} from only the low redshift sources and keeping the value of matter density of the Universe $\Omega_m$ fixed. These methods associate {a} probability to {each} galaxy as {a possible} host of the dark sirens \cite{Soares-Santos:2019irc}, and is only effective up to low redshift when the number of galaxies is limited. However, if the method is applied to the high redshift sources, then the possible host along a particular direction of the sky is going to be large in number, and as a result, the method is not informative enough to choose the correct galaxy as a host. As a result, it restricts the use of dark sirens to low redshift even if accurate distance measurement is possible for sources at high redshift from the LIGO/Virgo design sensitivity \cite{Acernese_2014,Martynov:2016fzi}, and from the upcoming gravitational wave detectors such as \s{the} Kamioka Gravitational Wave Detector (KAGRA) \cite{Akutsu:2018axf}, LIGO-India \cite{Unnikrishnan:2013qwa}, Laser Interferometer Space Antenna (LISA) \cite{2017arXiv170200786A}, \s{the} Einstein Telescope (ET) \cite{Punturo:2010zz}, \s{the} Cosmic Explorer (CE) \cite{Reitze:2019iox}\s{,} and \s{the} TianQin observatory \cite{Luo:2015ght, Wang:2019ryf, Liu:2020eko}. 
An alternative way to find the redshift of the source is by exploring any mass scale associated with the compact objects originating due to \s{the} neutron star mass distribution \cite{PhysRevLett.73.1878, PhysRevD.85.023535}, \s{the neutron star} tidal deformation \cite{Messenger:2011gi}\s{,} or using the mass-gap in the gravitational wave source population due to the pair-instability supernova \cite{Farr:2019twy}.

In this work, we explore a method that can be applied up to high redshift (up to which galaxy samples are going to be available) and can measure the value of $H_0$ along with the density of dark energy, \s{the} equation of state of dark energy, and also the spatial distribution of black holes with respect to the dark matter distribution. {The galaxy catalogs from the combination of several ongoing/upcoming surveys (such as SDSS/BOSS, \s{the} Dark Energy Survey (DES) \cite{10.1093/mnras/stw641}, \s{the} Dark Energy Spectroscopic Instrument (DESI) \cite{Aghamousa:2016zmz}, Euclid \cite{2010arXiv1001.0061R}, \s{the} Nancy Grace Roman Telescope \footnote{Previously known as Wide-Field InfraRed Survey Telescope \href{https://wfirst.gsfc.nasa.gov}{(WFIRST)}} \cite{2012arXiv1208.4012G, 2013arXiv1305.5425S, Dore:2018smn}, \s{the} Rubin Observatory \cite{2009arXiv0912.0201L}, Spectro-Photometer for the History of the Universe, Epoch of Reionization, and Ices Explorer (SPHEREx) \cite{Dore:2018kgp}) will be available up to redshift $z=3$. \s{The combination of different missions will be able to cover nearly the full sky}.}
We exploit the fact that both the gravitational wave sources and galaxies are
tracers of \s{the} matter density, and therefore\s{,} they are spatially correlated through the underlying matter field, to infer the redshift\s{s} of dark standard sirens \cite{1986Natur.323..310S, PhysRevD.93.083511, Mukherjee:2018ebj}. We build on previous work, where clustering with galaxies was applied to redshift unknown (or photometrically known) SNe \cite{Mukherjee:2018ebj}. Our method does not identify \textit{the host galaxy} of the BBH source but finds \textit{its host redshift shell} by exploring the three-dimensional spatial cross-correlation of the gravitational wave sources with redshift-known galaxies. Host galaxy identification is\s{,} therefore\s{,} at the limit of our approach that only exploits very small, galaxy-scale correlations \cite{Mukherjee:2018ebj}. The exploitation of the clustering aspect is also implemented to identify the redshift distribution of the galaxies \cite{Newman:2008mb, Menard:2013aaa, Schmidt:2013sba}. 

We detail the formalism of this method and the likelihood setup in Sec. \ref{formalism} and Sec. \ref{likelihood}\s{,} respectively. Our method does not require making any additional assumption about the redshift dependence of the merger rate of gravitational waves sources but only requires that the BBH mergers trace galaxies {(incorporating the possibility of natal birth kicks)}, so that there is a spatial correlation, as discussed in Sec. \ref{sims}. We show a forecast for the accuracy and precision of the measurements of {$H_0$} achievable with our method in Fig. \ref{allh0} after marginalizing over the matter density $\Omega_m$, and the redshift\s{-}dependent gravitational wave bias parameter $b_{GW} (z)= b_{GW}(1+z)^\alpha$. Details about this result are given in Sec. \ref{results}. Moreover, since dark sirens can be detected up to high redshift, this method also makes it possible to explore the expansion history of the Universe and provide an independent measurement of the cosmological parameters related to matter density $\Omega_m$, dark energy \s{equation-of-state} $w_0$, and its redshift dependence $w(z)= w_a(z/(1+z))$. This method can also explore the bias parameter of the gravitational wave sources at different redshifts $b_{GW}(z)$, which will capture its spatial distribution with respect to dark matter.
This method will also be applicable to the multi-messenger test of gravity proposed in \cite{Mukherjee:2019wcg, Mukherjee:2019wfw}. The breadth of the scientific returns possible from this avenue surpasses that of the statistical host identification methods \cite{Chen:2017rfc, Nair:2018ign, Soares-Santos:2019irc}. For comparison, we apply our method to only low redshift sources with a fixed value of $\Omega_m$, assuming a known value of the gravitational wave bias parameter $b_{GW}$. We find that the error-bar on $H_0$ from these methods \cite{Chen:2017rfc, Nair:2018ign, Soares-Santos:2019irc} is more by only about $30\%$ than our method. This implies \s{that} in the limit of low redshift sources, these methods \cite{Chen:2017rfc, Nair:2018ign, Soares-Santos:2019irc} approach the optimal solution proposed in this work. We conclude in Sec. \ref{conc}.

\section{Formalism: Exploring the clustering of the gravitational wave sources with galaxies}\label{formalism}
 {The matter distribution in the Universe at large scales ($\geq 100$ Mpc/h) is homogeneous and statistically isotropic, which agrees with the fundamental assumption known as Copernican principle \cite{Yadav:2005vv}. Under this setup, we can write the large scale distribution of galaxies in terms of a galaxy density field $\delta_g(\vec r)= n_g(\vec r)/\bar{n_g} -1$, where $n_g(\vec r)$ {is the number density of galaxies} at a position $\vec{r}$ and  $\bar n_g$ is the mean number density of galaxies\s{.}\footnote{$\bar n_g \equiv N_g/V_s= \sum_i\, n_g(\vec r_i)$} According to the standard model of cosmology, the spatial distribution of galaxies should trace the underlying distribution of matter in the Universe, and can be expressed as a biased tracer of the matter density field $\delta_{m}(\vec k)$ by the relation,}
\begin{equation}\label{deltag}
  \delta_g (\vec k)= b_g(k) \delta_{m} (\vec k),
\end{equation}
where $b_g(k)$ is the galaxy bias, $\delta_g (\vec k)$ is the Fourier transformation of the real space galaxy density field $\delta_g(\vec r)$. {The galaxy bias parameter encodes how galaxies trace the dark matter distribution \cite{Desjacques:2016bnm}.} 
 \s{ A spectroscopic (or photometric) survey observes galaxies in the redshift space denoted by {superscript} $s$.}
 {The motion of the galaxies along the line of sight induces a distortion in the position of the galaxies in redshift space, which causes anisotropy in the statistical properties of the observed density field, known as redshift space distortion (RSD)} \cite{1987MNRAS.227....1K, Hamilton:1997zq}. On large scale\s{s}, the effect is captured by the Kaiser term \cite{1987MNRAS.227....1K, Hamilton:1997zq}
\begin{align}
\label{deltagrsd}
\delta^s_g (\vec k, z)= b_g(k,z)(1+ \beta_g \mu_{\hat k}^2)\delta^r_{m} (\vec k, z), 
\end{align}
where $\beta\equiv f/b_g(k,z)$ \s{is} defined in terms of $f\equiv \frac{d\ln D}{d\ln a}$ which is the logarithmic derivative of the growth function $D$ with respect to the scale factor $a$, $\mu_{\hat k}= \cos{\hat n.\hat k}$ is angle between the line of sight and the Fourier mode $\hat k$, {and the superscript $r$ denotes real space}. {The growth factor $D= \frac{H(z)}{H_0}\int_z^\infty \frac{dz'(1+z')}{H(z')^3}[\int_0^\infty\frac{dz''(1+z'')}{H(z'')^3}]^{-1}$ captures the growth of the cosmological perturbations with redshift \cite{1980lssu.book.....P}.}
 
 Astrophysical gravitational wave events are expected to occur in galaxies, and \s{therefore, will} follow the spatial distribution of the galaxies with a bias parameter $b_{GW}$ that is different from the bias parameters for galaxies $b_{g}$\footnote{If primordial black holes (PBHs) are dark matter, then the distribution of PBHs {is} also going to be a biased tracer of the galaxy distribution.}. Following the definition Eq.~\eqref{deltag}, we can define the density field for the gravitational wave sources in real space $\delta^r_{GW}$ as
\begin{align}
\delta^r_{GW} (\vec k, z)= b_{GW}(k,z)\delta^r_{m}(\vec k, z),
\end{align}
where $b_{GW} (k, z)$ is the gravitational wave bias parameter \cite{Mukherjee:2018ebj, Mukherjee:2019qmm, Mukherjee:2019oma, Calore:2020bpd, Vijaykumar:2020pzn}. The gravitational wave bias parameter captures how gravitational wave sources trace the large scale structure in the Universe \cite{Mukherjee:2019oma}. Since the gravitational wave sources are tracers of luminosity distance and not redshift, they are not affected by RSD.

{The matter distribution also exhibits clustering property\s{,} which can be statistically described by the correlation function $\xi(r)$\footnote{Correlation function $\xi(r)$\s{,} is related to the power spectrum $P(k)$ by Fourier Transformation.} \cite{1975ApJ...196....1P, 1977ApJS...34..425D, 1983ApJ...267..465D, 1993ApJ...417...19H, 1993ApJ...412...64L}.}  {The spatial clustering of the multiple tracers such as galaxies and gravitational wave sources can be written in the Fourier space in terms of the three dimensional auto-power spectrum and cross power spectrum at different redshifts $z$ as\footnote{ {The angular bracket $\langle .\rangle$ denotes the ensemble average, which for a homogeneous and statistically isotropic Universe can be changed into an average over the spatial volume by ergodic theorem \cite{1980lssu.book.....P}.}}}
\begin{widetext}
\begin{align}\label{power-spec}
\begin{split}
\left\langle\begin{pmatrix}
  \delta^s_g (\vec k, z)\\
  \delta^r_{GW}(\vec k, z)
\end{pmatrix}\begin{pmatrix}
  \delta^s_g (\vec k', z)& 
  \delta^r_{GW}(\vec k', z)
\end{pmatrix}\right\rangle=
  \begin{pmatrix}
    P^{ss}_{gg} (\vec k, z)\delta_D( \vec k- \vec k') + \bar n_g(z)^{-1}&
     P^{sr}_{g\,GW} (\vec k, z)\delta_D( \vec k- \vec k')\\
    P^{sr}_{g\,GW} (\vec k, z)\delta_D( \vec k- \vec k') &
     P^{rr}_{GW\,GW} (\vec k, z))\delta_D( \vec k- \vec k') + \bar n_{GW}(z)^{-1}
\end{pmatrix},
\end{split}
\end{align}
\end{widetext}
where $P^{ij}_{xy} (\vec k, z)$ denotes the three dimensional power spectrum at redshift $z$ associated with the clustering between two tracers ($\{x\,y\} \in \{g, GW\}$) in redshift space or real space ($\{i\,j\} \in \{s, r\}$), $\delta_D( \vec k- \vec k')$ denotes the Dirac delta function, and $\bar n_{x}(z)^{-1}$ is the shot noise contribution which is non-zero only when $x$ and $y$ are the same. {The shot noise contribution arises due to the discrete sampling of galaxies or gravitational waves sources.}  
The redshift tomographic estimate of the auto power spectrum ($x=y$) and cross power spectrum ($x\neq y$) between galaxies and gravitational wave sources can be written in terms of the matter power spectrum $P_m(k,z)$ as 
\begin{equation}\label{power-spec-dm}
\begin{split}
   P^{ss}_{gg} (\vec k,z)&= b^2_g(k,z)(1 + \beta_g \mu_{\hat k}^2)^2P_{m}(k,z),\\
   P^{sr}_{g\,GW} (\vec k,z)&= b_g(k,z)b_{GW}(k,z)(1 + \beta_g \mu_{\hat k}^2)P_{m}(k,z),\\
   P^{rr}_{GW\,GW} (\vec k,z)&= b^2_{GW}(k,z)P_{m}(k,z).
\end{split}
\end{equation}
 {\s{Astrophysical sources of gravitational waves are expected to form within galaxies. The latter, in turn, are expected to trace the underlying distribution of dark matter in the standard model of cosmology through the galaxy bias parameter }$b_g(k,z)$ \cite{1980lssu.book.....P, Desjacques:2016bnm}. As a result, astrophysical gravitational wave sources are also expected to trace the dark matter distribution, but with a different bias parameter $b_{GW}(k,z)$, which is yet to be measured. Even if a significant fraction of the gravitational wave sources would be of primordial origin (such as primordial black holes), they would still trace the underlying dark matter distribution because under gravitational instability they would cluster in regions of higher dark matter density. However, in such a scenario the gravitational wave bias parameter would exhibit a different behavior from the astrophysical gravitational wave bias parameter. The exploration of these differences is an interesting avenue to distinguish between astrophysical and primordial black holes. This will be studied in details in a future work \cite{Mukherjee:2020:new}.}

The presence of redshift space distortion (RSD) \cite{1987MNRAS.227....1K} induces anisotropy in the observed auto (and cross) power spectrum with galaxies, as shown in Eq. \eqref{power-spec-dm}. The bias parameter for galaxies $b_g(k,z)$ and gravitational wave sources $b_{GW}(k,z)$ \s{are redshift dependent and scale dependent}. At large scales ($k<0.1\,\rm{h/Mpc}$)\s{,} the galaxy bias is scale-independent, and behaves like a constant $b_g= 1.6$ \cite{2012MNRAS.427.3435A, Desjacques:2016bnm,2017MNRAS.470.2617A}. \s{We also expect similar scale-independent behavior of the bias parameter, $b_{GW}$ for gravitational wave sources, at} large scales ($k<0.1\, \rm{h/Mpc}$), {as it will be mainly affected by the large scale spatial distribution of the galaxies.} However, at smaller scales ($k>0.1\, \rm{h/Mpc}$), the gravitational wave bias parameter is likely to be scale-dependent {as it will depend on the cluster scale (and galaxy scale) astrophysical processes related to binary formation, stellar metallicity, and supernovae/AGN feedback}. The redshift dependence of the gravitational wave bias parameter is also unknown and we will discuss {its implication} in detail in the next section.

One of the key aspects of Eq.~\eqref{power-spec-dm} is that the underlying cross power spectrum between galaxies and gravitational wave sources $P^{sr}_{g\, GW}(\vec k, z)$ is related to the matter power spectrum $P_{m}(k, z)$, which is also measurable from the auto power spectrum of galaxies $P^{ss}_{gg}(\vec k, z)$. As a result, $P_{g\,GW}(\vec k, z)$ should follow similar statistical properties as $P^{ss}_{gg}(\vec k, z)$. We exploit this very simple model to use the spatial cross-correlation of galaxies with gravitational wave sources to infer the luminosity-distance--redshift relation\s{,} and hence the cosmological parameters.

\section{Likelihood for inferring the expansion history using dark standard sirens}\label{likelihood}

Let us consider a sample of $N_{GW}$ gravitational wave sources (denoted by $i$) for which we have inferred the luminosity distance $\{d^i_l\}$ to the source over a sky volume denoted by $V_s$. For each of these sources, there is also a measurement of the sky localization $\{\theta^i_{GW}, \, \phi^i_{GW}\}$ with a $68\%$ sky localization error $\Delta \Omega^i_{GW}$ for each source. {Within the area of the sky localization, the exact position of a gravitational wave source is not known. As a result, any spatial information about the gravitational wave source within that region is smoothed out. This results \s{in smoothing the density field} of the gravitational wave sources in Fourier space for the comoving modes $k> k_{\rm eff}(z)\equiv \sqrt{8\ln2}/(\Delta \Omega_{GW}^{1/2}d_c(z))$, where $d_c (z)$ is the comoving distance to the source.\footnote{Comoving distance $d_c (z)$ is related to the luminosity distance $d_l (z)$ by the relation $d_l(z)= (1+z) d_c(z)$.}} Critically, assuming a Gaussian distribution of the sky localization error, we can write the effect of sky localization on the density field as $\delta_{GW}(\vec k, \Delta \Omega_{GW}, z)= \delta_{GW}(\vec k, z)e^{-k^2/k^2_{\rm eff}(z)}$. Along with the gravitational wave sources, we consider a number of galaxy samples $N_g= \bar n_gV_s$ in the overlapping sky volume $V_s$ with the known redshift $z_g$ and an error $\sigma_z$ and the sky position denoted by $\{\theta_{g}, \, \phi_{g}\}$ with an error on the sky position $\Delta \Omega_{g}$.\footnote{For all practical \s{purposes}, sky localization error for galaxies can be considered to be zero.} Using galaxy samples with known redshift, we can make tomographic bins of the galaxies with $N_z$ galaxies in each redshift bin.  

 {At this point, we can combine the measurements to infer the underlying cosmology; in particular, the expansion history of the Universe, which we model as} 
$H(z)= H_0(\Omega_m(1+z)^3 + \Omega_{de}\exp{(3\int_0^z d\ln(1+z)(1+w(z))})^{0.5}$, 
and the corresponding cosmological parameters ($\Theta_c \in$ $\{H_0$, $\Omega_m$, $w(z)= w_0 + w_a(z/(1+z))\}$) can be explored from dark standard sirens using the Bayes theorem \cite{bayes}
\begin{widetext}
\begin{equation}\label{posterior-1}
\begin{split}
   \mathcal{P}(\Theta_c|\vec{\vartheta}_{GW}, \vec d_{g})\propto & \iint d\Theta_n\, dz \, \bigg[\prod_{i=1}^{N_{GW}}\, \, \mathcal{L}(\vec{\vartheta}_{GW}| P^{ss}_{gg}(\vec k,z), \Theta_n, \vec d_g(z)) \mathcal{P}(\vec d_g| P^{ss}_{gg}(\vec k,z))
   \mathcal{P}({\{d^i_l\}}_{GW}|z, \Theta_c, \{\theta^i,\, \phi^i\}_{GW})\Pi(z) \bigg] \\& \times\, \Pi(\Theta_n)\Pi(\Theta_c),
\end{split}
\end{equation}
\end{widetext}
where, the gravitational wave data vector is composed of $\vec{\vartheta}_{GW} \equiv \{ d^i_l,\, \theta^i_{GW},\, \phi^i_{GW}\}$ and the galaxy data vector is composed of $\vec d_{g} \equiv \{\delta_g( z_g^i,\,\theta^i_{g},\, \phi^i_{g})\}$. $\Pi(\Theta_c)$ and $\Pi(\Theta_n)$ denote the prior on the cosmological parameters $\Theta_c$ and prior on the nuisance parameters $\Theta_n \in \{b_{GW}(k,z)\}$\s{,} respectively. $\Pi(z)$ denotes the prior on the redshift range of the gravitational wave sources {which can be taken \s{to be} uniform over a wide range. In the presence of a redshift information about the gravitational wave sources, an informative prior on the redshift can be considered. In this analysis\s{,} we consider a uniform prior $\mathcal{U}(0,1)$\footnote{$\mathcal{U}(a,b)$ denotes the uniform function over the range (a,b).} on the redshift unknown gravitational wave sources. This is sufficiently wide \s{enough} for the near-term and medium-term gravitational wave surveys we are considering. }

$\mathcal{P}({\{d^i_l\}}_{GW}|z, \Theta_c)$ is the posterior on the luminosity distance $d_l$ from the gravitational wave data $\vec{\vartheta}_{GW}$ which, for convenience, we model as a Gaussian distribution.\footnote{While this posterior is likely to be non-Gaussian in practice, we make this assumption purely to construct a forecast that can be compared \s{with} other studies making similar assumptions.} 
 
\begin{align}\label{pos-2}
\begin{split}
  \mathcal{P}({\{d^i_l\}}_{GW}|z, & \Theta_c, \{\theta^i,\, \phi^i\}_{GW}) \\ &\propto  \exp\bigg(-{\frac{(d^i_l(\{\theta^i,\, \phi^i\}_{GW})- d_l(z, \Theta_c))^2}{2\sigma^2_{d_l}}}\bigg),
  \end{split}
\end{align}
where, $\sigma_{d_l}$ is the error on the luminosity distance, and $d_l(z, \Theta_c)= (1+z) \int_0^z \frac{c\, dz'}{H(z')}$ {is the model for the luminosity distance}. {The luminosity distance error is marginalised over the inclination angle. The posterior on the luminosity distance is expected to be non-Gaussian for individual sources. In the cross-correlation technique, we combine the luminosity distance posteriors from multiple gravitational wave measurements within a luminosity distance bin. As a result, the combined posterior on the luminosity distance from $N_{GW}(d_l)$ sources approaches a Gaussian distribution due to the central limit theorem. So the assumption of Gaussian posteriors will not impact the results significantly when $N_{GW} (d_l)$ is large. In any case, in the method proposed in this paper, \s{a} non-Gaussian posterior for individual gravitational wave sources can be trivially included in Eq. \eqref{posterior-1}.}

The posterior of the galaxy density field $\mathcal{P}(\vec d_g| P_{gg}(\vec k, z))$ given the galaxy power spectrum $P_{gg}(\vec k, z)$ can be written as
\begin{equation}\label{pos-3}
  \mathcal{P}(\vec d_g| P^{ss}_{gg}(\vec k,z)) \propto \exp\bigg(-{\frac{ \delta^{s}_g(\vec k, z) \delta^{s*}_g(\vec k, z)}{2(P^{ss}_{gg}(\vec k,z) + n_g(z)^{-1})}}\bigg),
\end{equation}
where $\delta^s_g(\vec k, z)= \int d^3\vec r\, \delta_g (\vec r) e^{i\vec k .\vec r}$ is the Fourier decomposition of the galaxy distribution. The first term in the denominator $P^{ss}_{gg}(\vec k,z)$ is the galaxy three dimensional power spectrum defined in Eq.~\eqref{power-spec}, and $n_g(z)= N_g(z)/V_s$ is the number density of galaxies in the redshift bin $z$. {Due to the non-linear structure formation, the Gaussian approximation of the field can break down \s{at small} scales. In our analysis, we have taken the galaxy mock \s{catalog} (discussed in Sec. \ref{sims}) which has the log-normal distribution of the galaxy density field ($\bar \delta (r)= \ln (1+\delta_g (r))$). The statistics of the large scale structure density field can be be described well by the log normal distribution\s{,} as shown from cosmological simulations \cite{Coles:1991if, 1995ApJ...443..479B, Colombi:1994sj} as well as from observations \cite{1934ApJ....79....8H, Wild:2004me,Clerkin:2016kyr}.} 

The likelihood term $\mathcal{L}(\vec{\vartheta}_{GW}| P_{gg}(\vec k,z), \Theta_n, \vec d_g(z))$ in Eq.~\eqref{posterior-1} is \s{given by}\s{,}
\begin{widetext}
\begin{align}\label{likeli-1}
\begin{split}
  \mathcal{L}(\vec{\vartheta}_{GW}| P^{ss}_{gg}(\vec k,z), \Theta_n,& \vec d_g(z)) \propto \\ & \exp\bigg(-\frac{V_s}{4\pi^2}\int k^2 dk \int d\mu_k{\frac{ \bigg(\hat P (\vec k, \Delta \Omega_{GW}) - b_g(k,z)b_{GW}(k, z)(1 + \beta_g \mu_{\hat k}^2)P_{m}(k,z)e^{-\frac{k^2}{k^2_{\rm eff}}}\bigg)^2}{2(P^{ss}_{gg}(\vec k,z) + n_g(z)^{-1})(P^{rr}_{GW\,GW}(\vec k,z) + n_{GW}(z)^{-1})}}\bigg),
  \end{split}
\end{align}
\end{widetext}
where $\hat P(\vec k, z)= \delta_{g}(\vec k, z)\delta_{GW}^*(\vec k,\Delta \Omega_{GW})$, $n_{GW}(z)= N_{GW}(d^i_l(z))/V_s$ is the number density of gravitational wave sources denoted in terms of the number of objects in the luminosity distance bin $N_{GW}(d^i_l(z))$, and $V_s$ denotes the total sky volume. {The first term in the numerator $\hat P(\vec k, z)$ denotes the observed cross-correlation signal between galaxies and gravitational wave sources, while the second term is the model of the expected \s{cross-correlation} power spectrum in redshift space in the presence of the anisotropic RSD, bias parameters, and limited sky resolution \s{$\Omega_{GW}$} of the gravitational wave sources. The bias parameters can be considered as nuisance parameters (or also cosmological parameters), and are marginalised over in our setup. \s{The term in the denominator denotes the covariance matrix of the cross-correlation power \s{spectrum, in which} the first term arises from the variance of the galaxy distribution, and the second \s{term from }the variance of the gravitational wave distribution.} The cross-correlation power spectrum is integrated over the sky volume denoted by $V_s$ and the Fourier wave number $k$. The total number of Fourier modes which contributes to the signal depends on the volume of the sky survey given by $N_m= k^2dkV_s/4\pi^2$. The integration in Eq.~\eqref{likeli-1} takes into account the {anisotropic} shape of the power spectrum by combining the contribution from $\mu_k= \cos\hat n.\hat k$ arising due to the RSD.} The likelihood is maximized for \s{the} set of cosmological parameters\s{,} \s{which} transforms the galaxy density field from redshift space to match or maximally correlate with the spatial distribution of gravitational wave sources.  

 In the limit $n_x(z)P_{x}(k, z)> 1$, the likelihood is in the cosmic variance limited regime, while $n_x(z)P_{x}(k, z)< 1$, it is in the shot noise dominated regime. For the gravitational wave sources expected within $5$ years (with an event rate $R(z)= \,100$ Gpc$^{-3}$ yr$^{-1}$ \cite{LIGOScientific:2018mvr, LIGOScientific:2018jsj}), we \s{will} explore the cross-correlation between the galaxies and gravitational wave sources only for small values of $k<k_{\rm eff}$ in the shot noise regime $n_{GW}P^{ss}_{GW\,GW}(k,z)<1$. Galaxy samples \s{will} have $\mathcal{O}(10^9)$ galaxies \cite{2009arXiv0912.0201L, 2010arXiv1001.0061R, 2012arXiv1208.4012G, 2013arXiv1305.5425S, Aghamousa:2016zmz, Dore:2018smn, Dore:2018kgp} and as a result, we will be in the cosmic variance limited regime for the values of $k<k_{\rm eff}$. So the denominator of the exponent in Eq.~\eqref{likeli-1}, is going to scale as $ \frac{4\pi^2P^{ss}_{gg}(\vec k,z)}{n_{GW}(z)}$. With the availability of large numbers of gravitational wave samples, the measurement \s{will be} in the cosmic variance limited regime $n_{GW}P^{rr}_{GW\,GW}(k,z)>1$, \s{in which} case the denominator of the exponent can be approximated as $4\pi^2P^{ss}_{gg} (\vec k,z) P^{rr}_{GW\,GW}(\vec k,z)$. In this analysis, we have considered an analytical covariance matrix. This can also be calculated from simulations for a specific mission of large scale structure and gravitational waves experiment.

\section{Generation of mock catalog}\label{sims}
We implement our method on a mock catalog of large scale structure and gravitational wave sources which are produced {for the log-normal distribution of the density field using the publicly available package} \texttt{nbodykit} \cite{Hand:2017pqn}. {The realization of the galaxies and gravitational wave sources are obtained from the same random realization, using a fixed matter power spectrum $P_m(\vec k, z)$ with different bias parameters for galaxies and gravitational wave sources $b_g$ and $b_{GW}$ respectively.} {As the cosmic density field evolves with redshift \cite{1980lssu.book.....P}, we need to take this into account in our cosmological simulations. To achieve this, we have generated several mock catalogs with box size (in units of Mpc/h) $[l_x=1350,\, l_y=1350,\, l_z= 300]$ at each redshift bin, starting from $z=0$ to $z=1.0$ with Planck-2015 cosmology \cite{Ade:2015xua}. Then all these \s{mock catalogs} are combined to obtain a single mock catalog over the entire redshift range. The galaxy distribution and the gravitational wave sources are chosen from this distribution which already includes cosmological evolution as a function of redshift. The method is also repeated for finer/wider choices of the $l_z$ and the results obtained from our method are robust.} This \s{mock catalog} \s{does} not take into account the contribution from weak lensing, since it is going to have a marginal ($\leq 1\%$) increase in the variance of the {inferred cosmological parameters for the low redshift gravitational wave sources} considered in this analysis.

\textit{Galaxy samples:} The galaxy samples are produced for a scale-independent bias parameter $b_g=1.6$ including the effect from RSD \cite{Hand:2017pqn}. The galaxy mocks are obtained for the number of galaxies $N_g= 1.5\times 10^4$. The redshift of these sources is assumed to be known spectroscopically, which implies \s{that} the {corresponding} error in the redshift measurement is $\sigma_z \approx 0$.

\textit{Gravitational wave samples:} For the same set of cosmological parameters and \s{the} same realization of the large scale structure density field from which we produced the galaxy samples, we obtain the gravitational wave samples $N_{GW}$ \footnote{Different cases of $N_{GW}$ are considered in this analysis, and are discussed in the respective sections} with the gravitational wave bias parameter $b_{GW}(z)= b_{GW}(1+z)^{\alpha}$ with $b_{GW}=2$ and $\alpha= 0$. For these samples we consider three different cases for sky localization error ${\Delta \Omega_{GW}}= 10$ sq. deg., ${\Delta \Omega_{GW}}= 25$ sq. deg., and ${\Delta \Omega_{GW}}= 100$ sq. deg. \cite{Fairhurst:2010is, Chan:2018fpv} which are possible to achieve from the network of five gravitational wave detectors (LIGO-Hanford, LIGO-Livingston, Virgo, KAGRA, LIGO-India \cite{Unnikrishnan:2013qwa, Acernese_2014, Martynov:2016fzi, Akutsu:2018axf}). For each gravitational wave source, the fractional error on the luminosity distance depends inversely on the matched filtering signal-to-noise ratio ($\rho$) \s{given by the relation} \cite{Sathyaprakash:1991mt,Cutler:1994ys,Balasubramanian:1995bm, 2010ApJ...725..496N,Ghosh:2015jra}
\begin{equation}\label{snr}
  \rho^2\equiv 4\int_0^{f_{max}} df \frac{ |h(f)|^2}{S_n(f)},
\end{equation}
 where the value of $f_{max}$ is considered as $f_{merg}= c^3(a_1\eta^2 + a_2\eta +a_3)/\pi G M$ \cite{Ajith:2007kx} \footnote{$M= m_1+m_2$ is the total mass of the coalescing binary, $\eta$ is the symmetric mass ratio $\eta= m_1m_2/M^2$, $c$ is the speed of light and $G$ denotes the gravitational constant. The values of the parameters are $a_1= 0.29740$, $a_2=0.044810$, $a_3=0.095560$ \cite{Ajith:2007kx}.}, $S_n(f)$ is the detector noise power spectrum, which we consider as the advanced LIGO design sensitivity \cite{Martynov:2016fzi} \footnote{The noise curves are available publicly on this website \href{https://dcc-lho.ligo.org/LIGO-T2000012/public}{https://dcc-lho.ligo.org/LIGO-T2000012/public}}.  
 The template of the gravitational wave strain $h(f)$ for $f\leq f_{merg}$ can be written in terms of the redshifted chirp mass $\mathcal{M}_z= (1+z)\mathcal{M}_c$, inclination angle with respect to the orbital angular momentum $\hat L.\hat n$ (which is denoted by the function $\mathcal{I}_{\pm} (\hat L.\hat n)$), and luminosity distance to the source $d_L$ by the relation \cite{1987thyg.book.....H, Cutler:1994ys,Poisson:1995ef,maggiore2008gravitational, Ajith:2007kx}
 \begin{equation}\label{strain}
   h_{\pm}(f)= \sqrt{\frac{5}{96}}\frac{G^{5/6}\mathcal{M}_z^2 (f_z\mathcal{M}_z)^{-7/6}}{c^{3/2}\pi^{2/3}d_L}\mathcal{I}_{\pm} (\hat L.\hat n).
 \end{equation}
In this analysis, we critically consider the posterior distribution of luminosity distance to be Gaussian with the minimum matched filtering detection threshold $\rho_{th}=10$ for equal mass binaries with masses $30\, M_{\odot}.$\footnote{$M_\odot= 2\times 10^{30}$ kg denotes the mass of the sun.} The fractional error in the luminosity distance $\sigma_{d_l}/d_l$ can be about $10\%$ for the bright sources having high detection SNR $\rho>60$ and as large as $70\%$ for the objects at detection threshold $\rho=10$. {Sources with poor sky localization and large error on the luminosity distance will contribute only \s{a} marginal improvement to the estimation of the cosmological parameters. So only the fraction of events with better sky localization will effectively improve the precision and accuracy of the estimation of the cosmological parameters. With the network of GW detectors such as LIGO-Hanford, LIGO-Livingston, Virgo, KAGRA (and in the future from LIGO-India), a sky localization error less than 100 sq. deg. is achievable \cite{Chan:2018fpv}.}
The mean values of the luminosity distance are set to those of a flat LCDM cosmological model with parameter values $[H_0=\, 70\,\text{km/s/Mpc},\,\Omega_m= 0.315,\,\Omega_\Lambda= 1-\Omega_m, w_0=-1, w_a=0]$. The {chosen value of the Hubble parameter, $H_0$ is} completely different from {that} considered in the large scale structure mock catalog ($H_0= 67.3$ km/s/Mpc) to show that the inferred cosmological parameters are affected only by the luminosity distance and not by the parameters assumed in the mock catalog.
For gravitational wave sources, we do not assume any redshift information. The current estimate of the event rate of BBHs is $R(z)= 10^2$ Gpc$^{-3}$ yr$^{-1}$ \cite{LIGOScientific:2018mvr}. With this event rate, we expect a few thousand \s{of events to be detected every year} with the advanced LIGO design sensitivity \cite{Martynov:2016fzi}. In this analysis, we show the measurability of the expansion history \s{by considering a} few different cases of the number of gravitational wave sources $N_{GW}$\footnote{We consider four cases of $N_{GW}= 50, 100, 200, 280$ for this analysis in the LIGO design sensitivity, which is expected to be easily available with the network of gravitational wave detectors.} and for the sky localization which is expected to be achievable with a network of four/five gravitational detectors. 

\begin{figure*}
\centering
\includegraphics[trim={0.cm 0.cm 0.cm 0.cm},clip,width=0.9\textwidth]{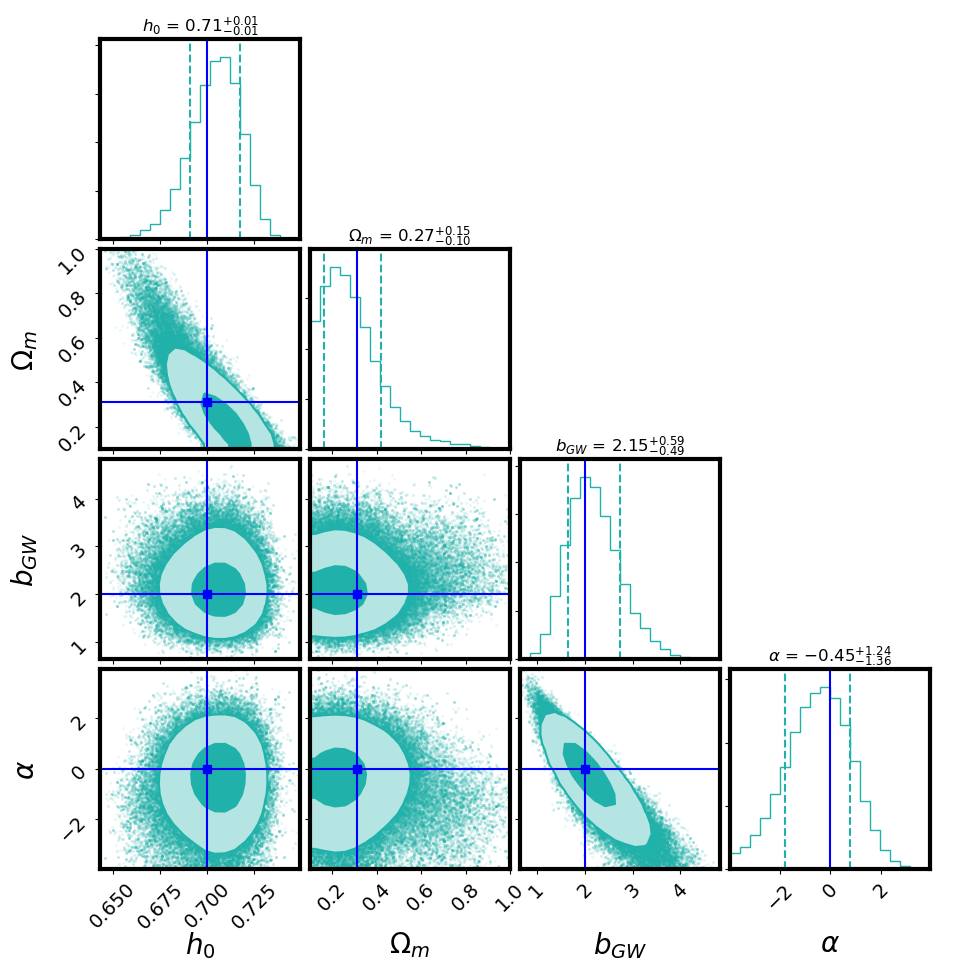}
\captionsetup{singlelinecheck=on,justification=raggedright}
\caption{We show the joint posterior of the cosmological parameters $H_0= 100h_0$ km/s/Mpc and $\Omega_m$ along with the nuisance parameters related to the gravitational wave bias parameter $b_{GW}(z)= b_{GW} (1+z)^\alpha$ for number of gravitational wave sources $N_{GW}(z)=40$ extended up to redshift $z=0.5$, and sky localization error $\Delta \Omega_{GW}= 10$ {sq.} deg. The $68\%$, and $95\%$ contours are shown in these plots along with the input values by the blue line. The mean value along with $1\sigma$ error-bar are mentioned in the title of the posterior distribution for all the parameters. Other cosmological parameters such as $w_0=-1$ and $w_a-0$ are kept fixed for these results.}
\label{hom}
\end{figure*}

\section{Results}\label{results}
Using the mock catalogs of galaxies and gravitational wave sources discussed in Sec. \ref{sims}, we explore the cosmological parameters which affects the expansion history of the Universe\footnote{Considering only the cosmological models with curvature $\Omega_K=0$.} (Hubble constant $H_0$, matter density $\Omega_m$, dark energy equation of state $w (z)$) using the formalism described in Sec.~\ref{likelihood}. The precise and accurate inference of the cosmological parameters using this method will rely on successfully mitigating the uncertainties associated with the unknown bias parameter and its redshift dependence associated with the gravitational wave sources. So\s{,} along with the cosmological parameters, we also consider the gravitational wave bias parameter \s{$b_{GW}(z)= b_{GW}(1+z)^\alpha$ to be unknown} and jointly infer the value of $b_{GW}$ and $\alpha$ (these are our nuisance parameters $\Theta_n \in\{b_{GW}, \alpha\}$) in the analysis along with the cosmological parameters. We consider three cases in this analysis: (i) $H_0$, $\Omega_m$, with fixed $w_0=-1$, and $w_a=0$; (ii) $\Omega_m$ and $\Omega_\Lambda$, with fixed $H_0=70$ km/s/Mpc, $w_0=-1$, and $w_a=0$; (iii) $w_0$ and $w_a$ with fixed $H_0=70$ km/s/Mpc and $\Omega_m=0.315$. Uniform priors on the cosmological and nuisance parameters are considered in the following range: $\Pi\bigg(\frac{H_0}{\text{km/s/Mpc}}\bigg) = \mathcal{U}(20, 150)$, $\Pi(\Omega_m) = \mathcal{U}(0.1,1)$, $\Pi(\Omega_\Lambda) = \mathcal{U}(0,1)$, $\Pi(w_0) = \mathcal{U}(-2,0)$, $\Pi(w_a) = \mathcal{U}(-8,8)$, $\Pi(b_{GW}) = \mathcal{U}(0,6)$, $\Pi(\alpha) = \mathcal{U}(-4,4)$ and $\Pi(z)= \mathcal{U}(0,1)$. We show the results only for the $\Delta \Omega_{GW}=10$ sq. deg. However, the results for $\Delta \Omega_{GW}=25$ sq. deg. only deteriorates marginally. For sky-localization error $\Delta \Omega_{GW}=100$ sq. deg., the impact on the error-bars are about a factor of two on the inferred parameters. {Even with the increase in the sky localization errors, our method still \s{gives unbiased} results for all the cosmological parameters and the bias parameters.}

\subsection{Measurement of $H_0$, $\Omega_m$ and $b_{GW}(z)$} The joint-estimation of the cosmological parameters $H_0$ and $\Omega_m$ along with the nuisance parameters are shown in Fig.~\ref{hom} for fixed value of $w_0=-1$ and $w_a=0$. These results are obtained for the cases with $N_g= 1.5\times 10^4$, $N_{GW}=200$ \footnote{The total number of gravitational wave sources $N_{GW}= \int N(z) dz$.}, and $\Delta \Omega_{GW}= 10$ sq. deg \footnote{Results with $\Delta \Omega_{GW}= 25$ sq. deg\s{.} changes only marginally.}. Results show that we can make the measurement of {$H_0= 70$ km/s/Mpc} with an accuracy of $1.9\%$ with only $N_{GW}(z)=40$ BBHs in each redshift bin of width $\Delta z=0.1$ up to redshift $z=0.5$ detectable with the advanced LIGO design sensitivity \cite{Martynov:2016fzi}. The \s{result shown} in Fig. \ref{hom} also indicates that the gravitational wave bias parameters $b_{GW}$ and $\alpha$ are uncorrelated with the cosmological parameters $H_0$ and $\Omega_m$. As a result, uncertainty associated with the gravitational wave bias parameter does not affect the inference of the cosmological parameters ({for the parametric form of the bias considered in this analysis}). This makes our method both precise and accurate to infer the cosmological parameters. Using this method we can measure the value of the gravitational wave bias parameter \s{with} $\sigma_{b_{GW}}/b_{GW}\sim 27\%$, with only $200$ BBHs at the advanced LIGO design sensitivity \cite{Martynov:2016fzi}. The cross-correlation technique makes it possible to measure the bias parameter even with the currently ongoing detector network and much before the operation of next-generation gravitational wave detectors \cite{Punturo:2010zz, Reitze:2019iox} by using the autocorrelation between the gravitational wave sources. This is another additional gain which is not possible from the other proposed methods \cite{Chen:2017rfc, Nair:2018ign, Soares-Santos:2019irc}.

\begin{figure*}
\centering
\includegraphics[trim={0.cm .cm .0cm .0cm},clip,width=.9\linewidth]{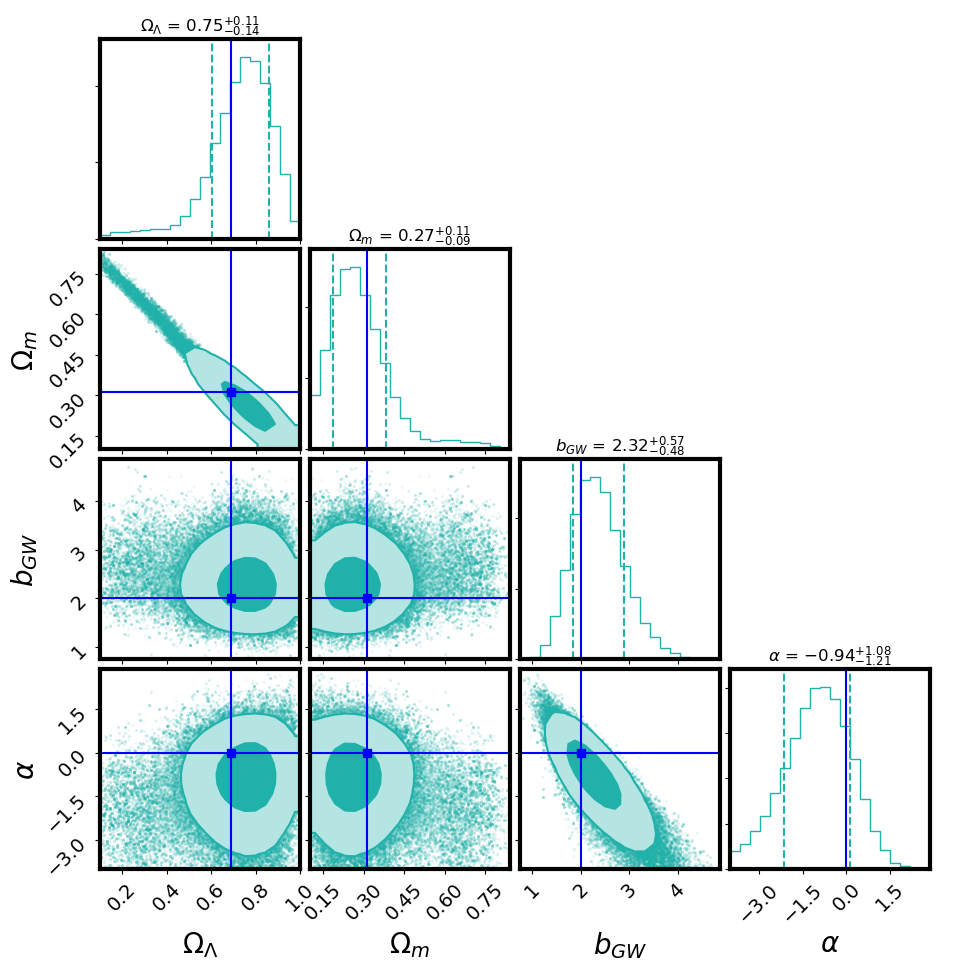}
\captionsetup{singlelinecheck=on,justification=raggedright}
\caption{We show the joint posterior of the cosmological parameters $\Omega_\Lambda$ and $\Omega_m$ along with the nuisance parameters related to the gravitational wave bias parameter $b_{GW}(z)= b_{GW} (1+z)^\alpha$ for number of gravitational wave sources $N_{GW}(z)=40$ extended up to redshift $z=0.7$, and sky localization error $\Delta \Omega_{GW}= 10$ {sq.} deg. The $68\%$, and $95\%$ contours are shown in these plots along with the input values by the blue line. The mean value along with $1\sigma$ error-bar are mentioned in the title of the posterior distribution for all the parameters. Other cosmological parameters such as $H_0=70$ km/s/Mpc, $w_0=-1$ and $w_a=0$ are kept fixed for these results. }\label{olamdaom}
\end{figure*}

The forecast posteriors on $H_0$ (after marginalizing over $\Omega_m, \,b_{GW},\, \alpha$) for $N_{GW}= 50,$ $100 $ and $200$ gravitational wave sources are shown in Fig.~\ref{allh0} along with the measurement of Hubble constant $H_0=67.4\pm{0.5}$ km/s/Mpc and $H_0=74\pm{1.4}$ km/s/Mpc from Planck \cite{Aghanim:2018eyx} and SH0ES \cite{Riess:2019cxk} respectively. 
The {uncertainty} in the measurement of $H_0$ decreases as the number of sources increases ($\sim N_{GW}^{-1/2}$) and as the uncertainty in the luminosity distances decreases ($\sim\sigma_{d_l}/d_l$). 

Fig.\ref{allh0} shows that a measurement of $H_0$ from 200 dark sirens ($\sigma_{H_0}/H_0=1.9\%$) compares favourably with that \s{which could be obtained from 50 sources} \textit{with EM counterparts} (such as BNS and NS-BH, assuming $\sigma_{H_0}/H_0=2\%$, \cite{Chen:2017rfc, Feeney:2018mkj}). However, as the number of detected dark sirens is expected to outnumber the sources with EM counterparts (such as BNSs and NS-BHs), one can expect the constraints on $H_0$ from dark sirens to dominate those from BNSs and NS-BHs, with very conservative assumptions about the availability of galaxy redshift survey covering a substantial fraction of the sky.
 \textit{In summary, our method will provide both accurate and precise measurements of $H_0$ from dark sirens along with $\Omega_m$, and redshift dependent gravitational wave bias parameter $b_{GW}(z)$ from the network of the advanced (with \s{or} without optical squeezing) gravitational wave detectors. Combining these two independent constraints, \s{one} would achieve $\sigma_{H_0}/H_0\,\sim 1.4\%$, which is competitive with current constraints from standard candles \cite{Riess:2019cxk}.} \s{Following our work, a recent work \cite{Bera:2020jhx} \s{has} also obtained the constraints on $H_0$ for a fixed value of $\Omega_m$, and \s{ with constant gravitational wave bias parameter using the cross-correlation technique.}} 
 
\subsection{Measurement of $\Omega_\Lambda$, $\Omega_m$ and $b_{GW}(z)$}
As our method can be applied to high redshift (up to which galaxy surveys will be available), we can also measure the energy budget in dark energy $\Omega_\Lambda$ from dark sirens. We make the joint estimation of the cosmological parameters $\Omega_\Lambda$--$\Omega_m$ along with the two bias parameters $b_{GW}$ and $\alpha$ for the parametric form $b_{GW} (z)= b_{GW}(1+z)^\alpha$ for a fixed value of $H_0=70\,\text{km/s/Mpc}, w_0=-1$ and $w_a=0$ with $N_{GW} (z)= 40$ up to redshift $z=0.7$. The corresponding plot is shown in Fig. \ref{olamdaom}. We show for the first time that the energy budget of dark energy can be measured from using dark sirens detectable within the modest timescale with the advanced LIGO design sensitivity \cite{Martynov:2016fzi} with only $N_{GW}= 280$ BBHs. The $\Omega_m$ and $\Omega_\Lambda$ are also uncorrelated with the bias parameters ($b_{GW}$ and $\alpha$), and as a result will not affect the measurement of cosmological parameters. The measurement of $\Omega_\Lambda$ and $\Omega_m$ gets less \s{constrained} for \s{a} limited number of gravitational wave sources if the value of $H_0$ is not kept fixed. However, \s{a} joint estimation with $H_0$ is possible with more \s{number of} gravitational wave sources. This method is also useful for the future gravitational wave detectors such as LISA \cite{2017arXiv170200786A}, ET \cite{Punturo:2010zz}, CE \cite{Reitze:2019iox} and TianQin observatory \cite{Luo:2015ght} to measure $\Omega_\Lambda$, $\Omega_m$, and the gravitational wave bias parameter $b_{GW}(z)$. 
\begin{figure*}
\includegraphics[trim={0.cm 0.cm 0.cm 0.cm},clip,width=.9\linewidth]{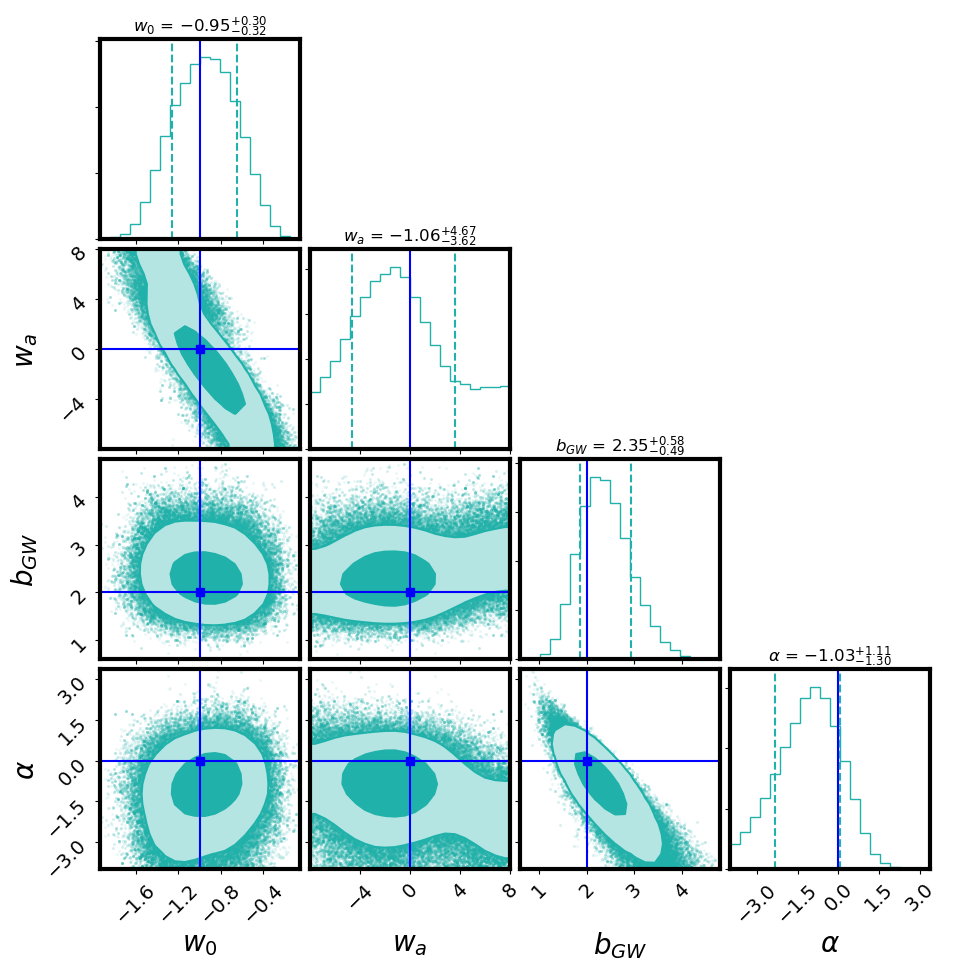}
\captionsetup{singlelinecheck=on,justification=raggedright}
 \captionsetup{singlelinecheck=on,justification=raggedright}
\caption{We show the joint posterior of the cosmological parameters $w_0$ and $w_a$ along with the nuisance parameters related to the gravitational wave bias parameter $b_{GW}(z)= b_{GW} (1+z)^\alpha$ for number of gravitational wave sources $N_{GW}(z)=40$ extended up to redshift $z=0.7$, and sky localization error $\Delta \Omega_{GW}= 10$ \s{sq.} deg. The $68\%$, and $95\%$ contours are shown in these plots along with the input values by the blue line. The mean value along with $1\sigma$ error-bar are mentioned in the title of the posterior distribution for all the parameters. Other cosmological parameters such as $H_0=70$ km/s/Mpc and $\Omega_m=0.315$ are kept fixed for these results.}
\label{w0wa}
\end{figure*}

\subsection{Measurement of $w_0$, $w_a$ and $b_{GW}(z)$}
The two-parameter phenomenological model of the dark energy equation of state $w_{de}= w_0 + w_a\,z/(1+z)$ is usually considered to explore the redshift dependence of dark energy. Using our method, we show the joint estimation of $w_0$ and $w_a$ along with the two bias parameters $b_{GW}$ and $\alpha$ (for the parametric form $b_{GW} (z)= b_{GW}(1+z)^\alpha$) in Fig. \ref{w0wa} for $N_{GW} (z)= 40$ extended up to $z=0.7$. We have kept the value of $H_0=70$ km/s/Mpc and $\Omega_m=0.315$ fixed for flat LCDM model. This plot shows that this technique is capable to infer the dark energy equation of state with $N_g=1.5\times 10^4$, $N_{GW}=280$ (up to redshift $z=0.7$) \s{and} for $\Delta \Omega_{GW}=10$ sq. deg. The constraints on the values on $w_0=-1$ are possible with $3.4\sigma$. \s{However, the} constraints on $w_a$ are going to be weak with the modest number of gravitational wave sources. With more number of gravitational wave sources possible from the five years of observation with the Advanced LIGO design sensitivity \cite{Martynov:2016fzi}, we will be able to infer the dark energy equation of state with higher accuracy (the error on the parameter reduces by $N_{GW}^{-1/2}$) for sources up to redshift $z\sim 1$.
{This independent avenue to measure $w_0$ and $w_a$ will also be} accessible from the next-generation gravitational wave detectors such as LISA \cite{2017arXiv170200786A}, ET \cite{Punturo:2010zz}, CE \cite{Reitze:2019iox} and TianQin observatory \cite{Luo:2015ght} for sources which are beyond redshift $z=1$. The gravitational wave bias parameters $b_{GW}$ and $\alpha$ are also uncorrelated with the parameters describing the dark energy equation of state and can be measured with high statistical significance as shown in Fig. \ref{w0wa}. {This method can also be used for the joint estimation of all cosmological parameters \emph{and} the gravitational wave bias parameter ($H_0, \Omega_m, \Omega_{de}, w_0, w_a, b_{GW}(k,z)$), provided one has a large number of gravitational-wave sources up to high redshift ($z>1$), so that the high redshift tomographic bins can be used to constrain $\Omega_m$, and the low redshift tomographic bins can be used to constrain the other cosmological parameters. This will be possible only from the next generation of ground-based detectors (such as ET, and CE), and also from the space-based detectors (such as LISA and TianQin observatory).}

\section{Conclusions and discussions}\label{conc}
Gravitational-wave sources are accurate luminosity distance tracers without requiring any external {astrophysical} calibration if instrument calibration can be achieved \cite{Sun:2020wke, Bhattacharjee:2020yxe}. {\s{An} accurate instrument calibration is essential for the measurement of the luminosity distance \cite{Sun:2020wke, Bhattacharjee:2020yxe}. If this is possible, then gravitational wave sources are an exquisite probe to measure the expansion history of the Universe by exploiting the luminosity distance and its redshift.} However, the inference of the redshift of the gravitational wave sources requires either an EM counterpart or a known mass scale (such as the mass scale associated with the tidal deformation \cite{Messenger:2011gi} \s{or} pair-instability of supernova \cite{Farr:2019twy}). For most of the gravitational wave sources, neither of these is available. {In this paper, we apply the method introduced in \cite{Mukherjee:2018ebj}, which exploits the scale associated with the three-dimensional clustering property of cosmic structure and the fact that both galaxies and gravitational wave sources, trace the underlying dark matter field} 
 
 Using the detector sensitivity expected from the current generation gravitational wave detectors \cite{Unnikrishnan:2013qwa, Acernese_2014, Martynov:2016fzi, Akutsu:2018axf}, we show that with a modest number of gravitational wave sources ($\sim 100$) we will be able to infer the Hubble constant with an accuracy $\sim2.5\%$ as shown in Fig. \ref{allh0} for gravitational wave sources distributed up to redshift $z=0.5$. The exploration of clustering of gravitational wave sources with galaxies makes it a robust method to infer the Hubble constant using the dark sirens. \s{Besides the Hubble constant measurement}, our method makes it possible to measure the fraction of dark energy in the Universe and its fundamental nature with the network of current generation gravitational wave detectors, as shown in Fig. \ref{olamdaom} and Fig. \ref{w0wa}. This is not possible currently from the gravitational wave sources with EM counterparts (such as BNS and NS-BH) due to the low observable horizon ($z<0.5$).  
 
 Along with the measurement of the expansion history, this method makes it possible to infer the gravitational wave bias parameter and its redshift dependence $b_{GW}(z)$. The gravitational wave bias parameter determines the spatial distribution of the gravitational wave sources with respect to the dark matter distribution and provides an avenue to measure this. Using our method, we can measure the bias parameter by more than $3\sigma$ precision with only $200$ BBHs distributed up to $z=0.5$, as shown in Fig.~\ref{hom}. With the availability of more gravitational wave sources, the bias parameter can be measured with higher precision and accuracy. The cross-correlation with the galaxies makes it possible to detect the bias parameters of gravitational wave sources sooner with higher statistical significance than \s{that} possible from the auto-correlation \cite{Vijaykumar:2020pzn}. 
 The redshift dependent bias parameter is not degenerate with the cosmological parameters as shown in Fig.~\ref{hom}, Fig. \ref{olamdaom}, and Fig. \ref{w0wa}, which makes it possible to reliably detect the cosmological parameters even if the gravitational wave bias parameter is currently unknown.  
 
 \s{On} longer timescale\s{s} corresponding to the launch of next-generation gravitational wave detectors such as LISA \cite{2017arXiv170200786A}, ET \cite{Punturo:2010zz}, and CE \cite{Reitze:2019iox}, we will be able to probe the expansion history of the Universe up to much higher redshift using the method proposed in this paper, without inferring the EM counterparts of the gravitational wave sources. So\s{,} the method proposed in this paper will help in building the observation strategy \s{for} the future gravitational wave detectors.
 
 {The success of the cross-correlation technique discussed in this paper depends on the availability of the EM telescopes with photometric/spectroscopic capabilities that can detect galaxies up to high redshift, with sufficiently low magnitude, and nearly full-sky coverage. Though such a rich data set is not possible from a single EM telescope, it is possible to achieve this by combining multiple EM telescopes operational in different EM frequency bands \s{ covering different sky areas.} Several ongoing \s{or} upcoming EM telescopes such as SDSS/BOSS \cite{Ross:2020lqz}, Dark Energy Survey (DES) \cite{10.1093/mnras/stw641}, Dark Energy Spectroscopic Instrument (DESI) \cite{Aghamousa:2016zmz}, Euclid \cite{2010arXiv1001.0061R}, Nancy Grace Roman Telescope \cite{2012arXiv1208.4012G, 2013arXiv1305.5425S, Dore:2018smn}, Rubin Observatory \cite{2009arXiv0912.0201L}, Spectro-Photometer for the History of the Universe, Epoch of Reionization, and Ices Explorer (SPHEREx) \cite{Dore:2018kgp}, \s{will play a key role in building up nearly full-sky galaxy catalogs which can be cross-correlated with the sources detectable from the network of gravitational wave detectors such as LIGO/ Virgo/ KAGRA.}} {Ground-based ongoing surveys such as DES will be measuring about 300 million galaxies over a sky area of approximately 5000 sq. deg up to redshift $z\sim 1.5$. In the future, SDSS-V will be measuring nearly full sky and about 2500 sq. deg with the Multi-Object Spectroscopy and Integral Field Spectroscopy over the redshift range $z=0.3$ to $z=2.5$. Also, surveys such as DESI and Rubin Observatory will be measuring galaxies up to $z=1.6$ and $z=3$ respectively with nearly about one-third of the sky area.} {Upcoming space-based missions such as Euclid, SPHEREx, and Nancy \s{Grace} Roman telescope will also play a vital role in the synergy with the \s{gravitational wave} detectors. Euclid will be measuring about 1.5 billion galaxies over a sky area of approximately 15000 sq. deg. in the redshift range $z=0.8-2$. SPHEREx will be mapping about 500 million galaxies over the full sky up to redshift $z\sim 2.5$. \s{The} Nancy \s{Grace} Roman telescope will be capable to measure about a few billion galaxies over a sky area of 2500 sq. deg. up to a redshift \s{of} $z=3$. \s{In a future work, we will explore the synergy between the EM telescopes and the next generation GW detectors such as ET \cite{Punturo:2010zz} and CE \cite{Reitze:2019iox} to map the cosmic history up to high redshift.}} 
 
 Finally, this method is not limited to gravitational wave sources but also applicable to any other distance tracers to infer the expansion history of the Universe using the luminosity distance -- redshift relation. Our method is readily applicable to SNe samples which will be detected with photometric redshift measurement from Rubin Observatory \cite{2009arXiv0912.0201L}, as already pointed \s{out} by a previous analysis \cite{Mukherjee:2018ebj}. In the future, this method can play a crucial role in \s{studying} cosmology with type-Ia SNe \cite{Scolnic:2019apa}. 
 This method will be useful in exploring the synergy between the upcoming missions such as DES \cite{10.1093/mnras/stw641}, DESI \cite{Aghamousa:2016zmz}, Euclid \cite{2010arXiv1001.0061R}, SPHEREx \cite{Dore:2018kgp}, Nancy Grace Roman Telescope \cite{2012arXiv1208.4012G, 2013arXiv1305.5425S, Dore:2018smn}. This method is also applicable to Fast Radio Burst (FRBs) \cite{Petroff:2019tty} to infer their redshift\s{s} for which host identification will be difficult.

\begin{acknowledgments}
The authors would like to thank Archisman Ghosh for carefully reviewing the manuscript and providing useful comments. S. M. also acknowledges useful discussion with Rahul Biswas, Neal Dalal, Will Farr, Archisman Ghosh, Salman Habib, Eiichiro Komatsu, Daniel M. Scolnic, Joseph Silk, and David Spergel. 
 This analysis was carried out at the Horizon cluster hosted by Institut d'Astrophysique de Paris. We thank Stephane Rouberol for smoothly running the Horizon cluster. SM and SMN are also supported by the research program Innovational Research Incentives Scheme (Vernieuwingsimpuls), which is financed by the Netherlands Organization for Scientific Research through the NWO VIDI Grant No. 639.042.612-Nissanke. This work is partly supported by the Delta ITP consortium, a program of the Netherlands Organisation for Scientific Research (NWO) that is funded by the Dutch Ministry of Education, Culture, and Science (OCW). The work of BDW is supported by the Labex ILP (reference ANR-10-LABX-63) part of the Idex SUPER, received financial state aid managed by the Agence Nationale de la Recherche, as part of the programme Investissements d'avenir under the reference ANR-11-IDEX-0004-02. The Center for Computational Astrophysics is supported by the Simons Foundation.  AS acknowledges support from the NWO and the Dutch Ministry of Education, Culture, and Science (OCW) (through NWO VIDI Grant No. 2019/ENW/00678104 and from the D-ITP consortium). In this analysis, we have used the following packages: Corner \cite{corner}, emcee: The MCMC Hammer \cite{2013PASP..125..306F}, IPython \cite{PER-GRA:2007}, Matplotlib \cite{Hunter:2007}, nbodykit \cite{Hand:2017pqn}, NumPy \cite{2011CSE....13b..22V}, and SciPy \cite{scipy}. The authors would like to thank the LIGO/Virgo scientific collaboration for providing the noise curves. LIGO is funded by the U.S. National Science Foundation. Virgo is funded by the French Centre National de Recherche Scientifique (CNRS), the Italian Istituto Nazionale della Fisica Nucleare (INFN), and the Dutch Nikhef, with contributions by Polish and Hungarian institutes.

\end{acknowledgments}

\def\urlprefix{}
\def\url#1{}
\bibliography{main_v3.bib}
\bibliographystyle{apsrev}
\end{document}